\pdfoutput=1
\documentclass{JINST}
\usepackage[utf8]{inputenc}
\usepackage{amsmath}
\usepackage{graphicx}
\usepackage{url}
\usepackage[mathscr]{eucal}

\usepackage{siunitx}
\usepackage{placeins}

\title{}

\title{First principles studies of the surface and opto-electronic properties of ultra-thin \textit{t}-Se}
\author{S. K. Barman$^a$, M. N. Huda$^a$, J. Asaadi$^a$, E. Gramellini$^b$, D. Nygren$^a$ \\ 
\llap{$^a$} University of Texas at Arlington, \\ 
701 S Nedderman Dr, Arlington, TX 76019, United States of America \\
\llap{$^b$}Fermi National Accelerator Laboratory, \\ 
            Chicago, Il 60616, United States of America \\}

\abstract{Selenium is an important  earth-abundant and non-toxic semiconductor with numerous applications across the semiconductor industries. Selenium has drawn attention from scientific communities  for  photovoltaics and imaging devices. Its usage as a photosensitive material largely involves the synthesis of the amorphous phase (a-Se) via various experimental techniques. However, the ground state crystalline phase of this material, known as the trigonal selenium (\textit{t}-Se), has not been extensively studied for its optimum electronic and optical properties. In this work, we present systematic studies based on density functional theory (DFT) for  ultra-thin $(10\overline{1}0)$ surface slabs of \textit{t}-Se. We report the surface energies, work function, electronic and optical properties as a function of number of layers for $(10\overline{1}0)$ surface slabs to access its suitability for applications as a photosensitive material.}


\begin{document}

\maketitle
\section{Introduction}\label{sec:Intro}
Selenium (Se) is a well known photovoltaic material, with its first studies dating back to the late 1800's \cite{OldSelenium, OldSelenium2}.  Commonly found in metal sulfide ores \cite{10.1579/0044-7447(2007)36[94:SGAH]2.0.CO;2}, Se is an earth-abundant material with a low melting point of $\sim 220^o$C, a relatively simple and stable crystalline structure, and  a relatively large direct band-gap \cite{PhysRev.158.623}. There are a few allotropes of Se, among them the thermodynamically stable triagonal Se (\textit{t}-Se)\footnote{sometimes referred to as hexagonal or crystalline Se}.  \textit{t}-Se has been observed to have a large optical absorption coefficient across a wide range of wavelengths.  A large absorption coefficient and a low melting point are  attractive properties that facilitate  the fabrication of large scale amorphous Se (a-Se) photovoltaic devices through vacuum deposition techniques.

Amorphous selenium has been the subject of research firstly for its application in xerographic photocopiers \cite{doi:10.1063/1.1702669,IEEEexample:Xerox}, then as a medium for direct X-ray imaging \cite{doi:10.1063/1.881994, doi.org/10.1023/A:1008993813689, Masuzawa_2013, doi:10.1118/1.2008428}, as well as for applications in composite films used in photocells \cite{Nakada_1985, Ito_1982, Todorov2017}. More recently research into Se's potential application as a general use photon detector with a potential wide wavelength application from the visible to UV \cite{https://doi.org/10.1002/pssr.201370438, 6399582}; see \cite{C8TC05873C} for a useful summary.

A new application of Se as a photon detector is foreseen  in gas and liquid based noble element particle detectors. Noble element detectors have become ubiquitous in  searches for dark matter and as a medium for detection of neutrino-nucleus interactions \cite{app11062455, BAUDIS201450}. The detection of the scintillation light  arising from  particle interactions is a key component. The  scintillation light for the most common noble element detectors (Argon \& Xenon) sits in the vacuum ultraviolet (VUV) spectrum (128nm – 175nm) or (9.7 eV – 7.1 eV). Conventional photon detectors such as photo-multiplier tubes (PMT’s) and multi-pixel photon counters (MPPC’s) have diminished  quantum efficiency at these wavelengths prompting both alternative detection techniques utilizing wavelength shifting coatings \cite{instruments5010004} and the development of new VUV sensitive MPPC’s \cite{Ieki:2018pbf}. Each of these approaches comes with their own complications, from the unintended behavior of the wavelength shifters \cite{Asaadi:2018ixs, Abraham:2021otn} to the small active surface areas of the individual VUV sensitive detectors \cite{OOTANI2015220}. In this landscape, techniques utilizing thin films of photosensitive semiconductors offers an opportunity for the development of alternative photon detector designs using new materials, such as a-Se. 

Some experimental work measuring the properties of Se based photon detector in the UV spectrum has taken place \cite{Leiga:68, Leiga:682,Masuzawa_2013, https://doi.org/10.1002/pssr.201370438, 6399582}, but little theoretical work modeling and understanding the opto-electronic properties of Se exists for photons in this frequency range. This work serves as a starting point for such calculations to better understand the properties that a selenium based detector might have for photons in the VUV spectrum. We begin with calculations for \textit{t}-Se modeling ultra-thin films which can be configured as fully relaxed free-standing layers. From these calculations, the surface energies can show the stability of ultra-thin layers of \textit{t}-Se and the density of states (DOS), band structures,  band gaps, and work function. These  can illuminate the optoelectronic properties, such as absorption coefficient, reflectivity, refractive index, and electron loss function for Se in the wavelength range most relevant for noble element detectors. In future works, the transition from t-Se to a-Se structures will be enabled by the theoretical calculations presented here. 

In this work, we present density functional theory based first principles studies of \textit{t}-Se $(10\overline{1}0)$ surface slabs having up to 15 atomic layers (45 atoms per unit cell). The paper is structured into three sections, the first describes the method of simulation for \textit{t}-Se (bulk and surface) using density functional theory (DFT) and the details of computational, thermodynamic, and optical modeling. The next section overviews the most important results pertaining to the optoelectronic properties. We conclude by outlining the next phase of the work. A supplemental section is provided at the end to give additional numerical and graphical details where relevant.

\section{Methods}\label{sec:methods}
The theoretical framework of this work is based on the density functional theory (DFT) \cite{Hohenberg1964, Kohn1965} implemented with the projector augmented wave method  (PAW) \cite{PAW-PhysRevB.50.17953, Kresse-PAW-PhysRevB.59.1758} as in Vienna \textit{ab initio} simulation package (VASP) \cite{KRESSE199615, Kresse-PhysRevB.54.11169}. Even though local magnetic structures are not expected, all the calculations performed in this work with spin polarization to allow additional freedom in electronic relaxations. For crystal structures and slabs visualization, VESTA \cite{vesta-Momma:db5098} was used throughout the work.

\begin{figure}[htb]
    \centering
    \includegraphics[scale=0.6]{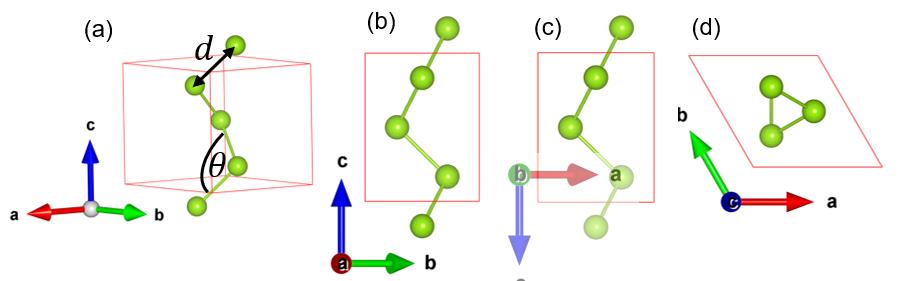}
    \caption{The crystal structure of \textit{t}-Se in (a) unit cell representation, which is viewed from (b) \textit{a}, (c) \textit{b}, and (d) \textit{c} lattice directions. The solid red line represents the boundary of a unit cell. Se atoms form a helical chain along \textit{c} lattice direction. \textit{d} is the distance between two consecutive Se atoms in the helical chain, and $\theta$ is the bond angle of Se atoms in that chain.}
    \label{fig:t-Se}
\end{figure}

Figure \ref{fig:t-Se} shows the crystal structure of \textit{t}-Se which contains three selenium atoms per unit cell. Figure \ref{fig:1x1slab} shows a typical slab geometry of stacked \textit{t}-Se used in the calculations.



\subsection{Computational Details}\label{subsec:DFT-VASP}

In this work, generalized gradient approximation (GGA) functional by Perdew–Burke–Ernserhof (PBE) \cite{GGA-PBE-PhysRevLett.77.3865, GGA-PBE-errata-PhysRevLett.78.1396} for the exchange-correlation energy is deployed. Since selenium atoms in \textit{t}-Se are layered along $[10\overline{1}0]$ and $[01\overline{1}0]$ directions, as shown in Figure \ref{fig:t-Se}, and the inter-layer distance is over 4 \r{A}, we have included van der Waals corrections (vdW) to GGA-DFT through zero damping DFT-D3 method of Grimme \cite{Grimme-D3-doi:10.1063/1.3382344}. The crystal structure of trigonal selenium with experimental lattice constants and bond lengths \cite{PhysRev.158.623, doi:10.1063/1.3382344, Wyckorff,  doi:10.1021/ic50054a037} and GGA and GGA+vdW optimized lattice parameters are summarized in Tables \ref{tab:LatticeParametersIntSe} and \ref{tab:BondLength}.

\begin{table}[!htb]
    \centering
    \begin{tabular}{|c|c|c|c|c|c|c|}
    \hline
    & \multicolumn{6}{|c|}{Lattice parameters for $\textit{t}$-Se} \\
    \hline
     & a (\r{A}) & b (\r{A}) & c (\r{A}) & $\alpha (^\circ)$ & $\beta (^\circ)$ & $\gamma (^\circ)$\\
    \hline
    Experiment \cite{ PhysRev.158.623, Wyckorff, doi:10.1021/ic50054a037} & 4.366 & 4.366 & 4.954 & 90 & 90 & 120 \\
    \hline
    GGA & 4.504 & 4.504 & 4.504 & 90 & 90 & 120 \\
    \hline
    GGA + vdW & 4.184 & 4.184 & 5.127 & 90 & 90 & 120 \\
    \hline
    \end{tabular}
    \caption{Experimental and DFT (GGA and GGA+vdW) optimized lattice parameters of \textit{t}-Se. The van der Waals corrections (vdW) were applied through the zero damping DFT-D3 method of Grimme \cite{Grimme-D3-doi:10.1063/1.3382344}. Since GGA overestimates lattice parameter, the inter layer separation increases by 0.138 \r{A}. However, with van der Waals corrections to GGA, the inter layer separation decreases by 0.182 \r{A} than the experimental lattice. On the other hand, the bond length increases up to 0.058 \r{A} due to the helical chain stretching along [0001] directions as tabulated in Table \ref{tab:BondLength}.}
    \label{tab:LatticeParametersIntSe}
\end{table}

\begin{table}[!htb]
    \centering
    \begin{tabular}{|c|c|c|}
    \hline
     & Bond length, \textit{d} (\r{A}) & Bond angle, $\theta (^\circ)$ \\
    \hline
    Experiment \cite{ PhysRev.158.623, Wyckorff, doi:10.1021/ic50054a037} & 2.373 & 103.07  \\
    \hline
    GGA & 2.405 & 103.62 \\
    \hline
    GGA + vdW & 2.431 & 103.98 \\
    \hline
    \end{tabular}
    \caption{Bond length, \textit{d} between two consecutive atoms in selenium helical chain and the bond angle, $\theta$  as depicted in Figure \ref{fig:t-Se}. }
    \label{tab:BondLength}
\end{table}

In the analysis of the simulation, we find that the inter-layer separation decreased after van der Waals corrections were applied due to enhanced long-range interactions. In addition, the cohesive energy per atom calculated by GGA + vdW (2.734 $eV$) for bulk \textit{t}-Se is 0.203 $eV$ lower than the GGA (2.937 $eV$) computed value. The cohesive energy per atom was calculated using the following formula:

\begin{equation}
E_c = \frac{N E_{\text{atom}}-E_{\text{bulk}}}{N}
\end{equation}

where  $E_{\text{bulk}}$ is the total energy of bulk selenium per unit cell, $N$ is the total number of atoms in a unit cell, and $E_{\text{atom}}$ is the energy of a single selenium atom in a cubic box of dimension 10 \r{A}. The experimental cohesive energy is 2.25 $eV$/atom \cite{Cohesive-PhysRevB.27.6296}, which indicate that GGA+vdW agrees better with the experimental value compared to the GGA-only calculation of cohesive energy. The binding between the Se atoms in the \textit{t}-Se crystals is significantly weaker when compared to the binding, for example, of Si (4.62 $eV$/atom) \cite{cohesive-si-PhysRevB.43.14248}.  \\

\begin{figure}[htb]
    \centering
    \includegraphics[scale=0.15]{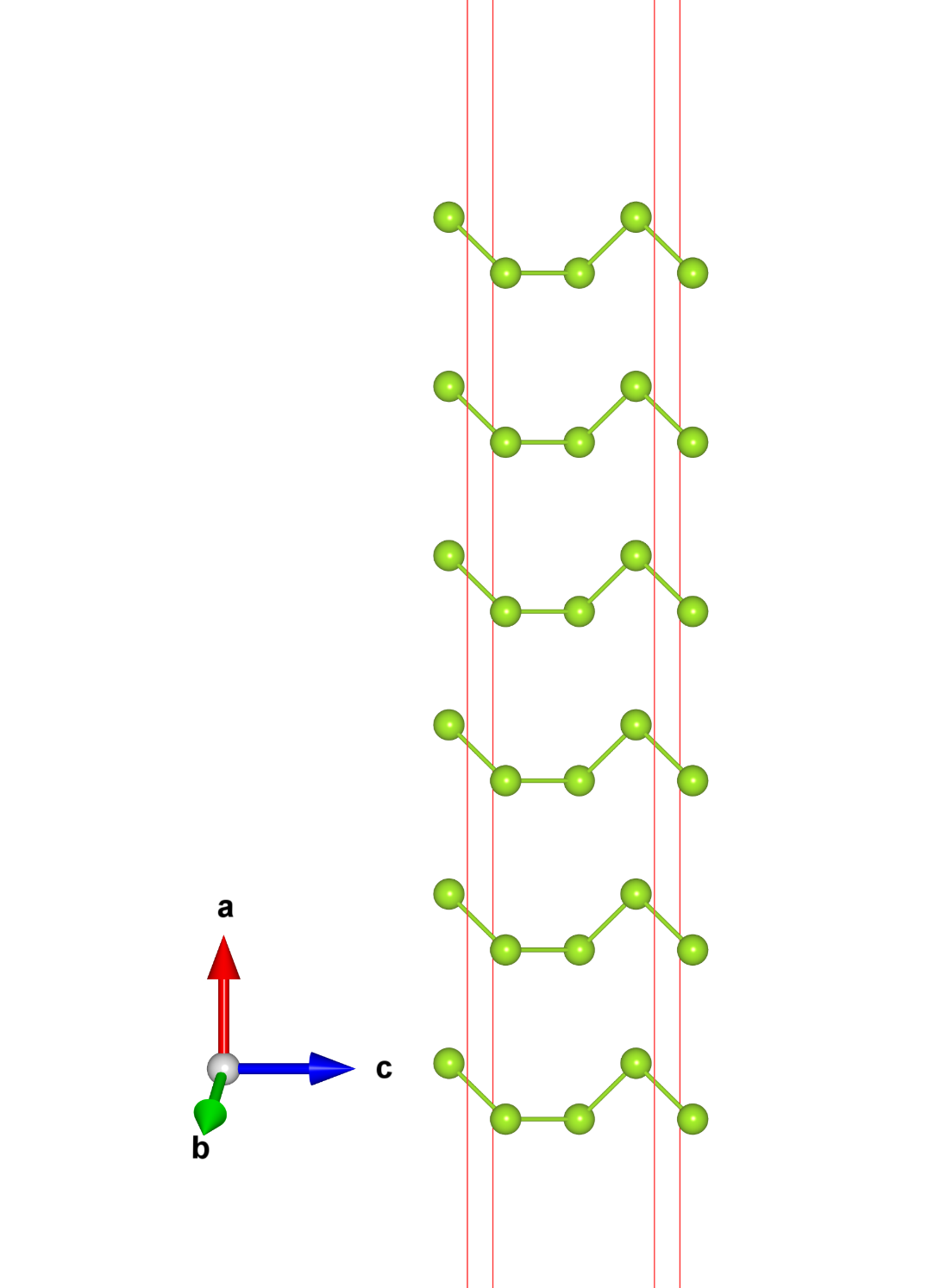}
    \caption{A schematic of a (1x1) slab geometry for stacked selenium atomic layers in $[10\overline{1}0]$ direction. Each simulation box that contains stacked selenium layers has a fixed  $25$ \r{A} vacuum along $[10\overline{1}0]$.}
    \label{fig:1x1slab}
\end{figure}

Since selenium atomic layers along $[10\overline{1}0]$ and $[01\overline{1}0]$ directions are arranged in the same geometric fashion in \textit{t}-Se crystals, atomic layers from a unit cell were stacked along $[10\overline{1}0]$ direction only with a 25 \r{A} vacuum for modeling slabs. A schematic of a six-layers $(1\times1)$ slab oriented in the $[10\overline{1}0]$ direction is presented in Figure \ref{fig:1x1slab}.


All slabs considered in this work for surface properties calculations were geometrically optimized without any symmetry constrained. Atoms in the simulation box could relax in all three lattice directions, and we name these relaxed layers as ``free-standing'' layers. Throughout the relaxation, cell volumes with vacuum layers (25 \r{A}) were kept fixed. The convergence criteria for the electronic self-consistent loop was set equal to or less than 1$\times$10$^{-8}$ \textit{eV}/\r{A}. All the ions were relaxed until the Hellman-Feynman force was equal to or less than 0.001 \r{A}. The kinetic energy cut off for the plane-wave basis set was 400 \textit{eV}. Se {4s$^{2}$ 4p$^{4}$} is the valence electrons configuration considered for this work. Each atomic layer in a (1$\times$1) slab contains three selenium atoms. In this work, the largest (1$\times$1) slab contains 15 atomic layers (45 atoms). The Brillouin zone was sampled by a 1$\times$21$\times$21 gamma centered \textit{k}-mesh for slabs having up to 10 layers (30 atoms) and by a 1$\times$15$\times$15 gamma centered \textit{k}-mesh for slabs containing more than 10 layers.

\subsection{Modeling of Surface Stability}\label{subsec:Thermodynamic-energy}

For the characterization of the stability of the $[10\overline{1}0]$ surface of \textit{t}-Se, the surface energy, $\gamma$ was calculated using the following formula:

\begin{equation}\label{eq:surface}
    \gamma = \frac{1}{2A}[E_\text{tot}(N) - N E_\text{Bulk}]
\end{equation}
where $E_\text{Bulk}$ is the total energy of bulk selenium per unit cell, $N$ is the number of layers, $E_\text{tot}(N)$ is the total energy of $N$-number of layers, and $A$ is the surface area. 

To avoid inconsistencies in the surface energies, $\gamma$ in Eq. (\ref{eq:surface}) resulting from a separate bulk calculation for $E_\text{Bulk}$ \cite{Boettger}, a well-established method is to derive $E_\text{Bulk}$ from a least square fitting of $E_\text{tot}(N) \sim N$ curve by re-writing the Eq. (\ref{eq:surface}) in the form :

\begin{equation}\label{eq:re-surface}
    E\text(N) = N E_\text{Bulk} + 2A\gamma
\end{equation}

The slope in Eq. (\ref{eq:re-surface}) provides the required $E_\text{Bulk}$ to calculate $\gamma$ from Eq.(\ref{eq:surface}). This method is tested throughly in previous works \cite{sajib-ga2o3-https://doi.org/10.1002/pssr.201800554, shafaq-MOTEN201637, edan-BAINGLASS2021121762}. In our case, the fitted values and separate bulk calculations for $E_\text{Bulk}$ are very close to one another as summarized in Table \ref{tab:TotalEnergy}. In particular, for both GGA and GGA + vdW  calculated energies, the difference is within 0.01 \textit{eV}. Details of the convergence for the surface energies and the least square fitting techniques are given in Figure \ref{fig:SurfaceAreaFree} and section \ref{sec:surfaceEnergy}.

\subsection{Modeling of Optical Properties}\label{Optical-formula}





In order to calculate the optical properties, we calculated the frequency-dependent complex dielectric function in the independent particle picture using VASP. The expression for the real ($\epsilon_1$) and imaginary ($\epsilon_2$) parts of the dielectric function ($\epsilon = \epsilon_1 + i\epsilon_2$) as a function of photon frequency ($\omega$) were as in ref.\cite{ALKHALDI2019e02908}. Using the following formulas, the optical absorption coefficient ($\alpha(\omega))$, reflectivity ($R(\omega)$), refractive index ($n(\omega))$ and the electron energy loss function ($L(\omega)$) were calculated for all the $(10\overline{1}0)$ surface slabs and bulk \textit{t}-Se:

\begin{equation}
    \alpha(\omega) = \frac{\sqrt{2}\omega}{c}\sqrt{ \sqrt{\epsilon_1(\omega)^2 + \epsilon_2(\omega)^2} - \epsilon_1 (\omega) }
\end{equation}

\begin{equation}
    R(\omega) = \left| \frac{\sqrt{\epsilon_1(\omega) + i\epsilon_2(\omega)} - 1}{\sqrt{\epsilon_1(\omega) + i\epsilon_2(\omega)} + 1} \right|^2 = \frac{(n-1)^2 + k^2}{(n+1)^2 + k^2}
\end{equation}

\begin{equation}
    n(\omega) = \sqrt{\frac{\sqrt{\epsilon_1(\omega)^2 + \epsilon_2(\omega)^2} + \epsilon_1 (\omega)}{2}}
\end{equation}

\begin{equation}
    L(\omega) = \frac{\epsilon_2 (\omega)}{\epsilon_1(\omega)^2 + \epsilon_2(\omega)^2}
\end{equation}

Where the extinction coefficient $k(\omega)$ is defined as

\begin{equation}
    k(\omega) = \sqrt{\frac{\sqrt{\epsilon_1(\omega)^2 + \epsilon_2(\omega)^2} - \epsilon_1 (\omega)}{2}}
\end{equation}


\section{Results and Discussion}\label{sec:Results}
The results of the simulation are discussed in four subsections. Section \ref{sec:surfaceEnergy} focuses on the stability analysis and reports the surface energies of $(10\overline{1}0)$ surface for \textit{t}-Se slabs containing up to 15 atomic layers.  Section \ref{sec:WorkFunction} shows the calculated work function to further understand the stability of those layers. Section \ref{sec:ElectronicProp} reports the calculated electronic properties such as band structures, band gaps and the density of states. Lastly, Section \ref{sec:OpticalProp} presents the calculated optical properties for all the relaxed layers considered in this work. As described above, all the calculations have considered the van der Waals corrections separately, along with GGA-DFT.

\subsection{Surface Energies}\label{sec:surfaceEnergy}

\begin{figure}[!htbp]
    \centering
    \includegraphics[scale=0.3]{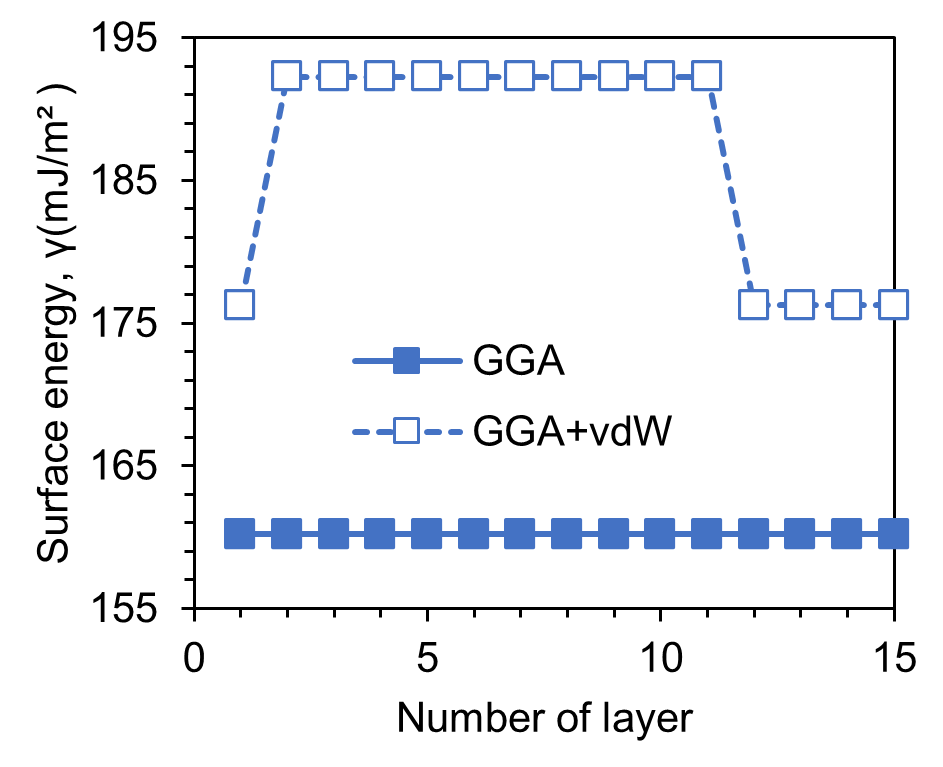}
    \caption{ GGA and GGA + vdW calculated surface energies $\gamma$ (mJ/m$^2$ ) as a function of the number of layers for fully relaxed \textit{t}-Se (100) surface slabs.}
    \label{fig:surfaceEnergies}
\end{figure}

The GGA calculated converged surface energy for free standing $(10\overline{1}0)$ \textit{t}-Se surface slabs is 160.22 $mJ/m^{2}$. When van der Waals interactions are added, the converged energy becomes approximately 16 $mJ/m^{2}$ higher, which is 176.24 $mJ/m^{2}$.  This increment is expected due to the long-range vdW interaction term added to the Kohn-Sham Hamiltonian. Our calculated values are in reasonably good agreement with experimentally measured surface energies of selenium (175 - 291 $mJ/m^{2}$) \cite{LEE1971213, Guisbeirs-doi:10.1063/1.4769358}. The surface energies shown in Figure \ref{fig:surfaceEnergies} are calculated values with respect to the number of layers in the slab model; numerical values are presented in Table \ref{tab:TotalEnergyByLayer} of Supplemental Material.  

\begin{figure}[!htbp]
    \centering
    \includegraphics[scale=0.26]{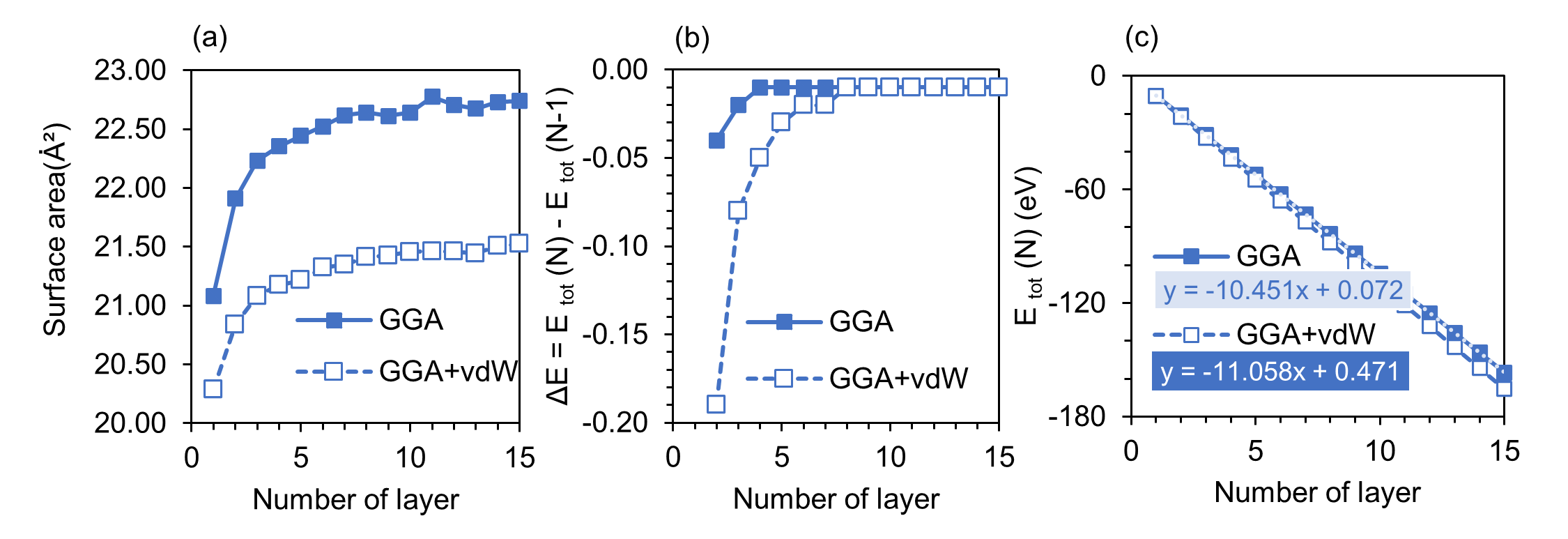}
    \caption{GGA and GGA + vdW optimized (a) surface area (\r{A}$^2$), (b) incremental energy $\Delta$E ($eV$) at each layer, and (c) energy E$_{\text{tot}}$ ($eV$) of free-standing \textit{t}-Se $(10\overline{1}0)$ surface slabs as a function of the number of layers (N). The figure in (a) shows that surface area converges within 0.01(0.02) \r{A}$^2$ for GGA(GGA +vdW) optimized layers while (b) shows the incremental energy, $\Delta$E convergence within 0.01 $eV$. Energies for bulk selenium, $E_{bulk}$ derived from least square fitting in (c) are very close to separate bulk calculations which are within 0.001 $eV$ per unit cell as in Table \ref{tab:TotalEnergy}.}
    \label{fig:SurfaceAreaFree}
\end{figure}

As a convergence criteria of surface energies calculation, the incremental energy per formula unit, or atomic layer, was calculated using the following formula 

\begin{equation}\label{eq:deltaE}
    \Delta E_\text{per formula unit} = E_{tot}(N+1) - E_{tot}(N),
\end{equation}
where $\Delta E_\text{per formula unit} \leq 0.01 eV$ (for both GGA and GGA + vdW). This is reported as a function of the number of layers, as shown in Figures \ref{fig:SurfaceAreaFree}(b).

 In addition, the total energies of layers are reported as a function of the number of layers, see Figures \ref{fig:SurfaceAreaFree}(c), to obtain the slope from a least square fitting to estimate for the energies of bulk selenium ($E_{Bulk}$).

Once $E_\text{Bulk}$ is found, $\gamma$ was calculated following Equation \ref{eq:surface} and were tabulated in Table \ref{tab:TotalEnergyByLayer}. From Figure \ref{fig:SurfaceAreaFree}, the layer-to-layer incremental energy saturates after seven layers for both GGA and GGA+vdW. However, the surface energy does not converge for GGA+vdW till twelve layers; there is a small fall in the value after eleven layers. The fall is due to the combined effect of larger surface area and higher layer-thickness per unit layer as the number of layers increased.

\begin{figure}[!htbp]
    \centering
    \includegraphics[scale=0.25]{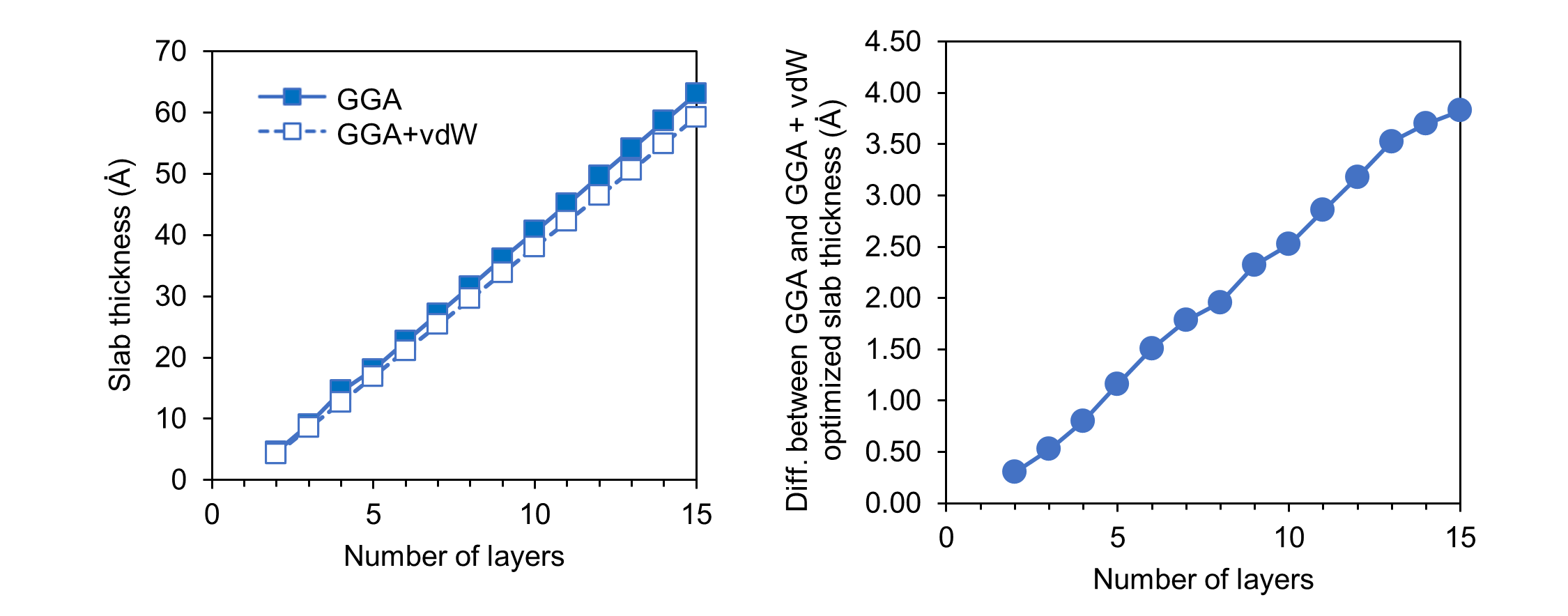}
    \caption{(Left) GGA and GGA + vdW optimized $(10\overline{1}0)$ surface slab thickness for \textit{t}-Se as a function of the number of layers. It is found that slab thickness increases almost linearly as the number of layers increases. However, the slab thickness for GGA + vdW optimized layers are less than GGA only optimized layers. For example, it is up to 3.82 \r{A} less for a 15 layers slab. (Right) Difference between GGA and GGA + vdW optimized slab thickness. As the number of the layers increases, the slab thickness difference also increases.  These results are summarized in Table \ref{tab:LayerThickness} in the Supplemental Material.}
    \label{fig:OptLayerThickness}
\end{figure}

The higher surface energy resulting from GGA + vdW calculations can also be explained from Figure  \ref{fig:OptLayerThickness} which shows that the rate at which the layer thickness increases with subsequent number of layers is higher for GGA calculated values than the GGA + vdW. This is due to the layer thickness for GGA + vdW-relaxed slabs being less and thus the inter-layer interaction increases, resulting in slightly higher surface energies.

\begin{table}[!htb]
    \centering
    \begin{tabular}{|c|c|c|}
    \hline
    & \multicolumn{2}{c|}{Total Energy (eV) of \textit{t}-Se per atomic layer or formula unit} \\
    \hline
    & \textbf{GGA} & \textbf{GGA + vdW} \\
    \hline
    Bulk & -10.450 & -11.059 \\
    \hline
    Least square fitting & -10.451 & -11.058 \\
    \hline
    \end{tabular}
    \caption{Total energy, $E_\text{bulk}$ (eV), per atomic layer for bulk \textit{t}-Se derived from separate bulk calculation and a least square fitting. Note that, each atomic layer or a unit cell contains three selenium atoms.}
    \label{tab:TotalEnergy}
\end{table}

\FloatBarrier
\newpage
\subsection{Work Function}\label{sec:WorkFunction}
\begin{figure}[!htbp]
    \centering
    \includegraphics[scale=0.25]{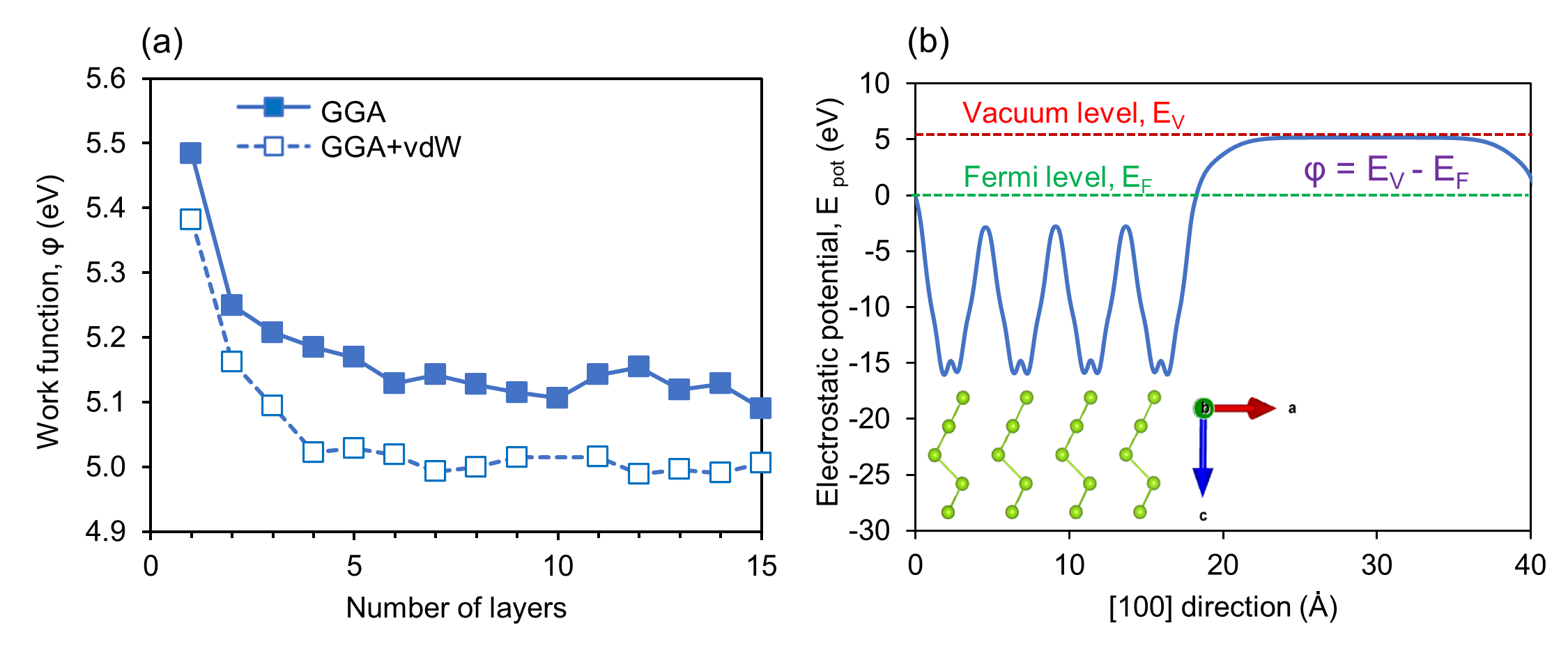}
    \caption{GGA and GGA + vdW calculated (a) work function $\phi$ (eV) of fully relaxed \textit{t}-Se $(10\overline{1}0)$ surface slabs up to 15 atomic layers. (b) a schematic of the electrostatic potential $E_{tot}$ ($eV$) for a four-layer slab calculated using the formula, $\phi = E_V - E_F$. Where $E_V$ is the vacuum level and $E_F$ is the Fermi level.}
    \label{fig:WorkFunction}
\end{figure}

To assess the surface stability further, we have calculated the work function of $(10\overline{1}0)$ surface of \textit{t}-Se. The work function of a semiconductor corresponds to the least amount of energy required to remove an electron from the bulk through the surface so that the electron no longer interacts with the material. It can be calculated as 

\begin{equation}\label{eq:work-function}
    \phi = E_V - E_F
\end{equation}

where $E_V$ is a vacuum level energy, a reference value within the slab model, and $E_F$ is the Fermi level. Within the slab model employed for the present study, the $E_V$ is defined as the saturated electrostatic potential energy as depicted in Figure \ref{fig:WorkFunction}(b). A higher work function indicates a higher stability of a surface, i.e., the surface is less reactive. The calculated work function for $(10\overline{1}0)$ surface slabs are presented in Figure \ref{fig:WorkFunction}(a) as a function of the number of layers. A single atomic layer of selenium has the highest work function (5.4 - 5.5 eV). As the number of layers increases, the work function decreases and tends to converge toward 5.1 eV for GGA optimized slabs. For GGA + vdW optimized slabs, this value is slightly lower but comparable to the bulk selenium surface work function (5.11 eV). Since minute changes in the surface configuration can change work function significantly, GGA + vdW optimized layers have lower values than the GGA relaxed layers. However, such higher values of work function for both cases (GGA and GGA+vdW) show good surface stability of the slabs considered for this work. Note, from four atomic layers, free-standing slabs have almost the same work functions as the bulk surface.

\subsection{Electronic Properties}\label{sec:ElectronicProp}
\begin{figure}[!htbp]
    \centering
    \includegraphics[scale=0.30]{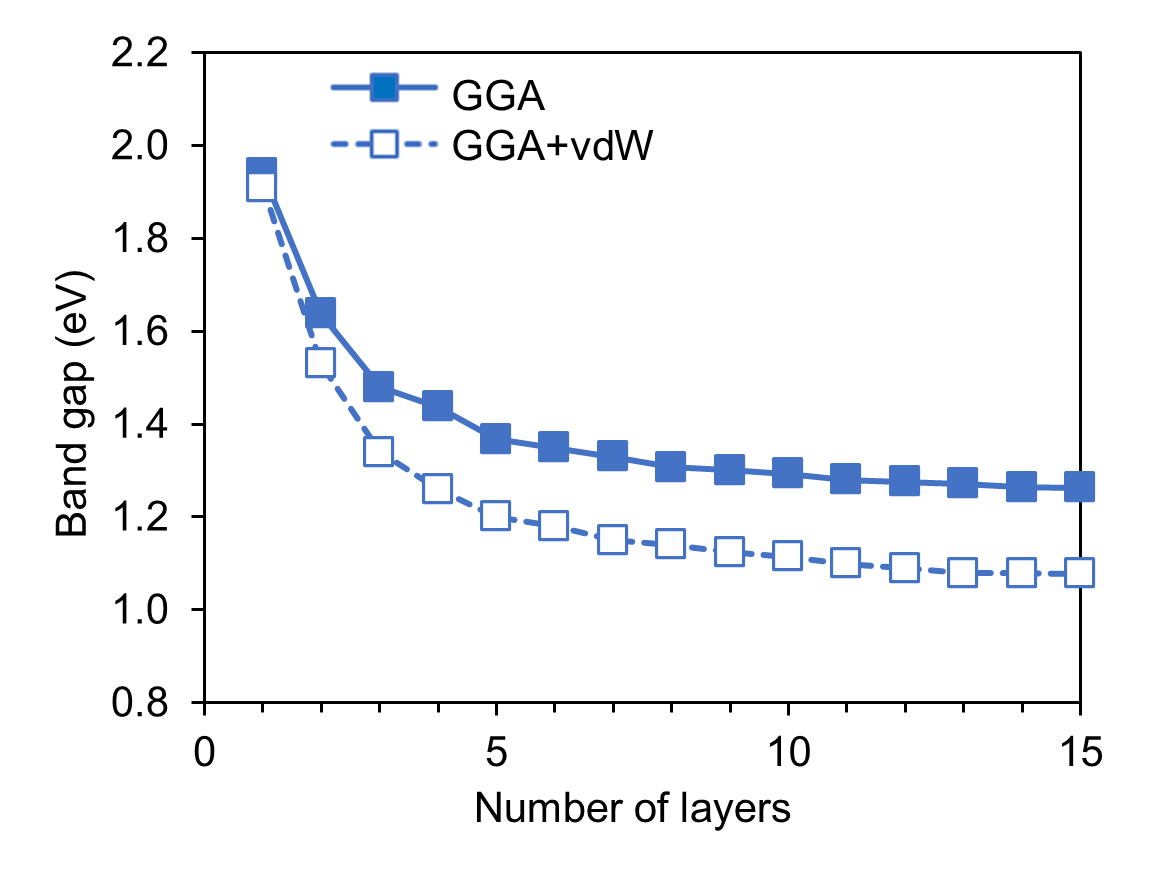}
    \caption{GGA and GGA + vdW calculated band gaps of fully relaxed \textit{t}-Se $(10\overline{1}0)$ surface slabs  as a function of the number of layers. }
    \label{fig:BandGap}
\end{figure}

Here the electronic properties of $(10\overline{1}0)$ surface \textit{t}-Se slabs are presented. We begin with the band gap which are calculated through GGA and GGA + vdW formalism. These are shown in Figure \ref{fig:BandGap}. We find that the band gap ($E_g$) is the highest for one atomic layer,  1.94 $eV$ from GGA and 1.91 $eV$ from GGA+vdW calculations, and much higher than the bulk band gaps (see Figure \ref{fig:BandStructure-Bulk} in Supplemental Material). As the number of layers increases, the band gap decreases and tends to reach the limit of the bulk selenium. The results can be justifies by the quantum confinement effect. Note, the calculated band gaps by either GGA or GGA+vdW method is an underestimation of the experimental band gap of 1.86 eV for \textit{t}-Se. \cite{Cohesive-PhysRevB.27.6296} The underestimation of band gaps is a well-known features by the standard DFT functionals. \cite{underestimation-gap-https://doi.org/10.1002/qua.560280846, underestimation-gap2-Perdew2801}

\begin{figure}[!htbp]
    \centering
    \includegraphics[scale=0.5]{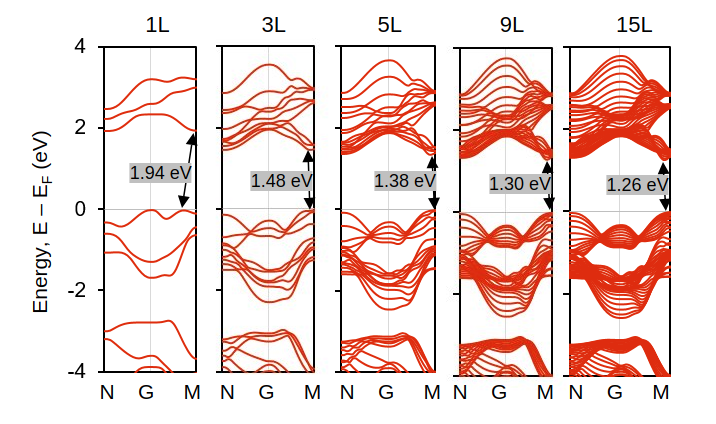}
    \caption{GGA calculated band structures for free-standing \textit{t}-Se $(10\overline{1}0)$ surface slabs containing one, three, five, nine and fifteen layers. The Fermi level is set to 0 eV. From corresponding density of states plots, it is found that the top of the valence band and the bottom of the conduction bands are predominantly contributed by the Se p orbitals. As the number of layers in the slab increases, the band gap decreases and tend to reach the band gap of the bulk \textit{t}-Se.}
    \label{fig:BandStructure}
\end{figure}

\begin{figure}[htb]
    \centering
    \includegraphics[scale=0.6]{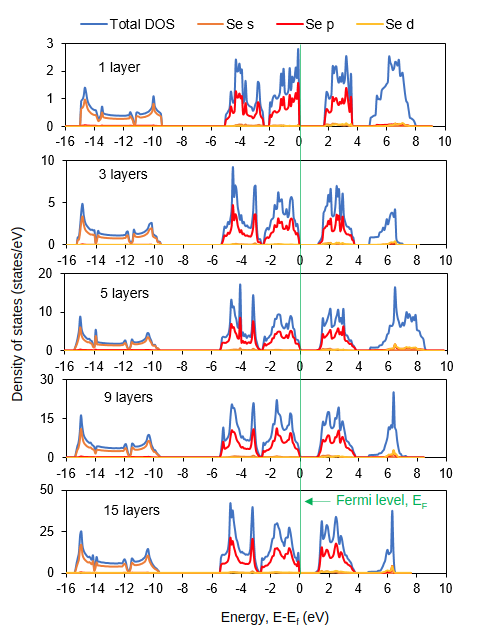}
    \caption{GGA calculated density of states for free-standing t-Se $(10\overline{1}0)$ surface slabs. The Fermi level is set to 0 eV. It is found that the top of the valence band and the bottom of the conduction bands are predominantly contributed by the Se p orbitals. As the number of layers in the slab increases, the band gap decreases and tend to reach the band gap of the bulk \textit{t}-Se.}
    \label{fig:DensityOfStates}
\end{figure}

Figures \ref{fig:BandStructure} and \ref{fig:DensityOfStates} show the GGA calculated band structure and density of states (DOS) respectively for the free-standing \textit{t}-Se surface slabs at one, three, five, nine, and fifteen layers. Analysis of these plots shows that the top of the valence band and the bottom of the conduction bands are dominated by the Se $p$ orbitals as like the case of the bulk Se. As the number of layers increases, the overall features of band composition and the density of states remain similar except the valence band maximum (VBM) shifting from G to M. However, the conduction band minimum (CBM) remains at the same point (near to M). There is an indirect to direct band gap transition as the number of layer increases. 


The calculations of the bulk \textit{t}-Se electronic properties including the direct and indirect band gaps and the corresponding DOS are shown in Section \ref{sec:Supp-Bulk-Elec-Prop} of the Supplemental Material in Figures \ref{fig:BandStructure-Bulk} and \ref{fig:DOS-Bulk}. Starting from few atomic layers, the DOS at the Fermi level and near CBM are very similar to that of the bulk. GGA + vdW calculated band structures and DOS are similar to the GGA calculated features with a slightly lower band gaps.

\FloatBarrier

\FloatBarrier

\subsection{Optical Properties}\label{sec:OpticalProp}

\begin{figure}[htb]
    \centering
    \includegraphics[scale=0.25]{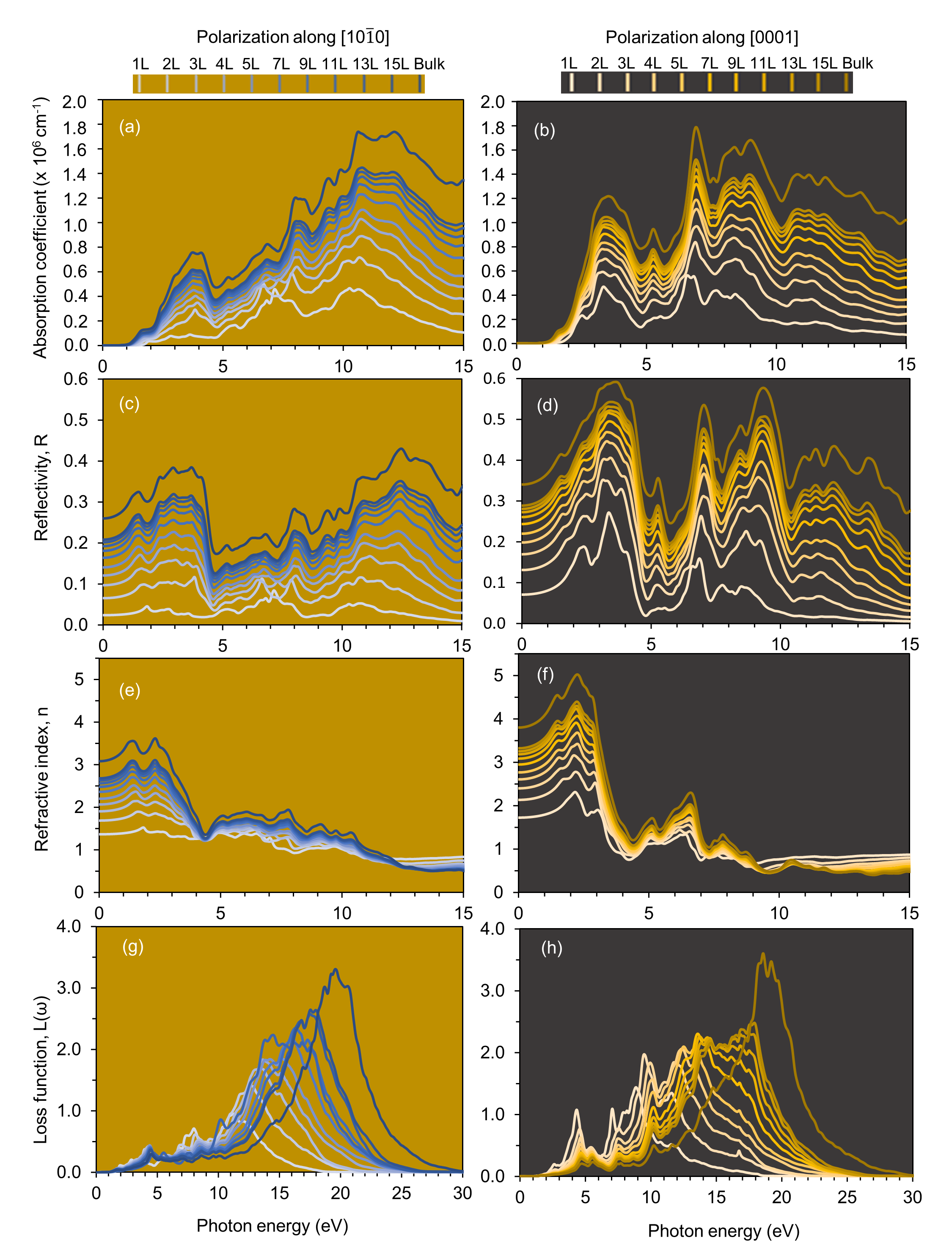}
    \caption{GGA calculated optical properties as a function of photon energy for free-standing \textit{t}-Se $(10\overline{1}0)$ surface slabs having up to fifteen atomic layers. Optical properties are calculated along $[10\overline{1}0]$ (left panel) and [0001] (right panel) directions.}
    \label{fig:OpticalProp-GGA}
\end{figure}

\begin{figure}[htb]
    \centering
    \includegraphics[scale=0.25]{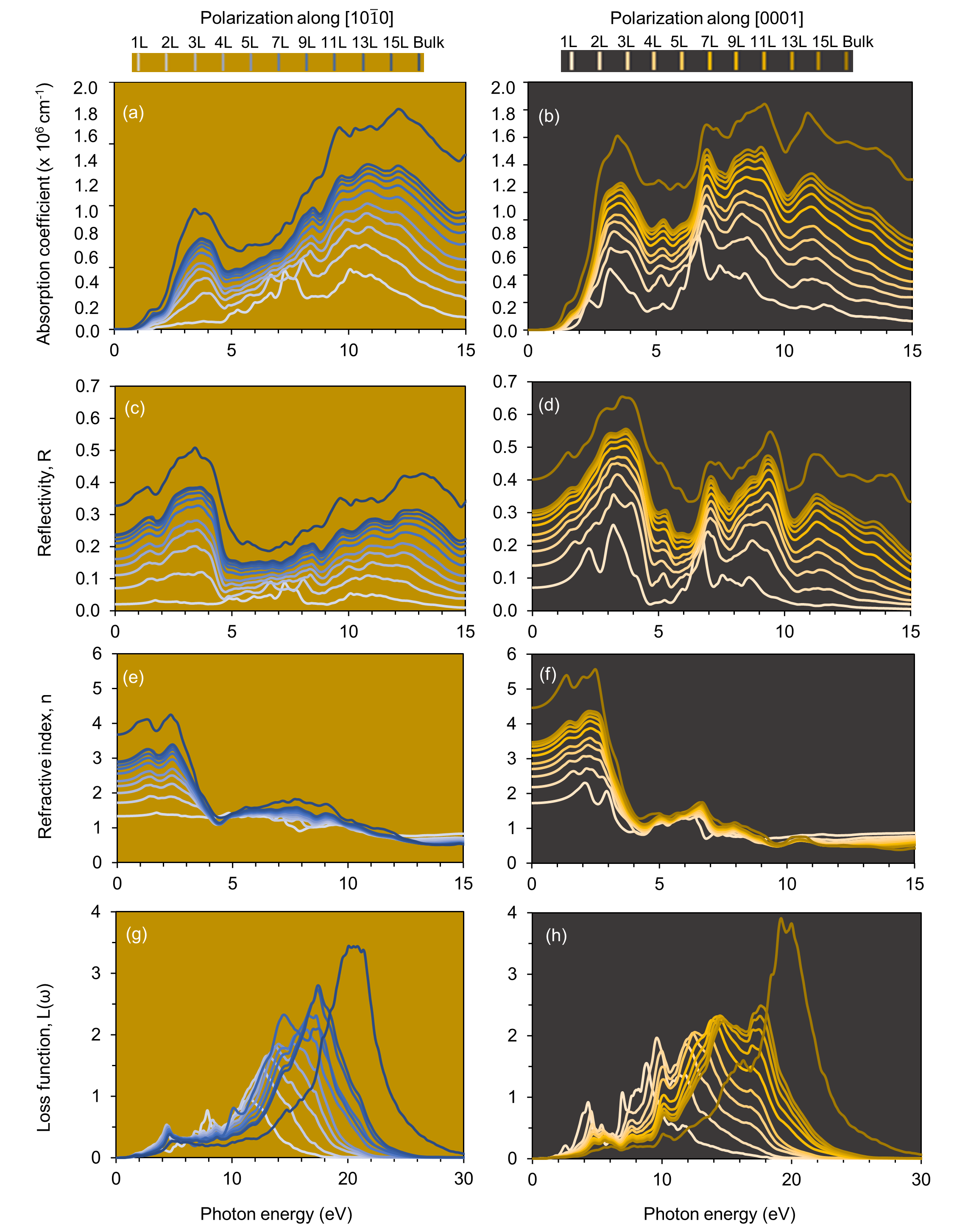}
    \caption{GGA + vdW calculated optical properties as a function of photon energy for free-standing \textit{t}-Se $(10\overline{1}0)$ surface slabs having up to fifteen atomic layers. Optical properties are calculated along $[10\overline{1}0]$ (left panel) and [0001] (right panel) directions.}
    \label{fig:OpticalProp-GGA+vdW}
\end{figure}

Using the analysis presented above, the optical properties of \textit{t}-Se are calculated using GGA and GGA+vdW across the photon energy range of interest and along the orientations $[10\overline{1}0]$ and [0001], as shown in Figures \ref{fig:OpticalProp-GGA} and \ref{fig:OpticalProp-GGA+vdW} for relaxed free standing layers. We focus our analysis on the behavior of the optical absorption coefficient ($\alpha$), the reflectivity (R), the refractive index (n), and the loss function (L).  

Some important observations which are apparent when comparing the optical absorption coefficient in the range relevant for the VUV spectrum of noble element detectors (e.g. 7.1 - 9.7 eV): i) after only a few atomic layers the behavior quickly begins to mimic the bulk properties suggesting that extremely thin-films will have a favorable absorption coefficient, ii) the absorption peaks near the area of our interest ($\sim$9 eV) in both polarizations and thus should have very favorable properties sought in their application as a VUV photodetector. 

\begin{figure}[htb]
    \centering
    \includegraphics[scale=0.26]{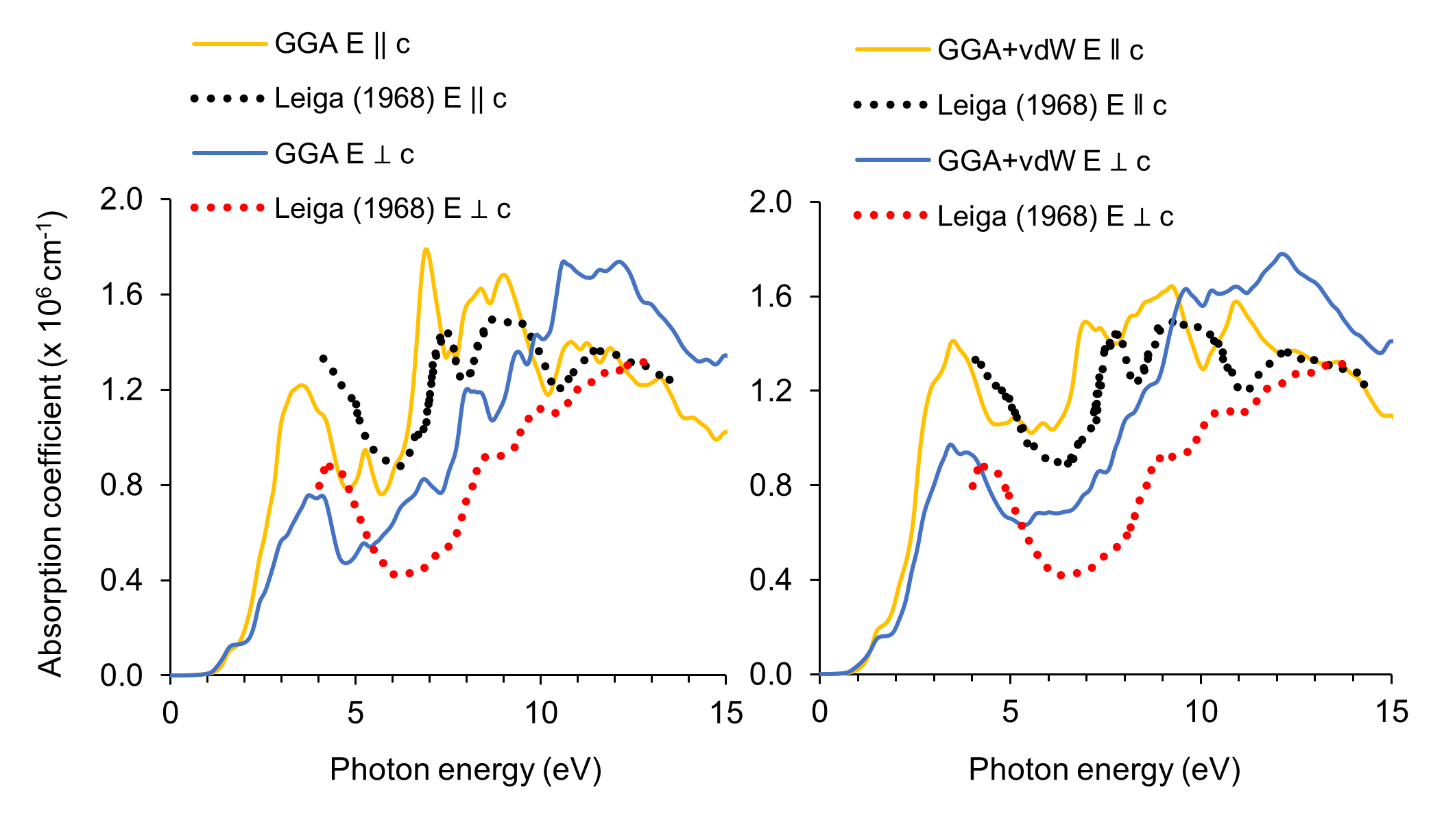}
    \caption{Using data taken by Leiga (1968) \cite{Leiga:68, Leiga:682} we can compare the calculations for bulk properties of the absorption coefficient ($\alpha$) made using GGA (Left) and GGA+vdW (Right) to experimental data taken when the photon is polarized parallel and perpendicular to the crystal. (\cite{Leiga:68}, \cite{Leiga:682})}
    \label{fig:AbsorptionComparison}
\end{figure}

Figure \ref{fig:AbsorptionComparison} shows the bulk calcuations compared to data from references \cite{Leiga:68} and \cite{Leiga:682} taken for $t$-Se when the incident radiation is parallel and perpindicular to the Se crystal over the photon energy range of 4-14 eV. The broad spectral features show general agreement while the two data sets differ from the GGA and GGA+vdW values by 10-20\% with the data being generally higher than the calculation for photon energies <5~eV and the calculations being higher than the data above 5~eV.

A similar analysis of the reflectance in Figure \ref{fig:OpticalProp-GGA} and \ref{fig:OpticalProp-GGA+vdW} shows: i) again, after only a few atomic layers the behavior quickly begins to mimic the bulk properties, ii)
the reflectivity is $\sim$30\% higher for the [0001] orientation compared to the $[10\overline{1}0]$ for both GGA and GGA+vdW, iii) the GGA+vdW calculation has a higher predicted reflectance in the region between 1-6 $e$, above which GGA and GGA+vdW agree with one another. Comparing our results to data a few conclusions can be drawn. Firstly, the broad spectral features show a similarity between the GGA calculation, GGA+vdW, and the data with peaks and drops occurring in roughly the same energies. In \textit{t}-Se the dip shown in the data around $\sim$6.5 eV and its rise thereafter is interpreted as a gap in the valence band. A similar, albeit, shifted phenomenon is seen in both the GGA and GGA+vdW calculation with the GGA+vdW more closely matching the data in the region of the dip. Secondly, the magnitude of the reflectance in the region of interest (7 eV - 9 eV) seems to be lower in the GGA calculation than the data by about $\sim$40\% and slightly less low in the GGA+vdW. 

\begin{figure}[htb]
    \centering
    \includegraphics[scale=0.26]{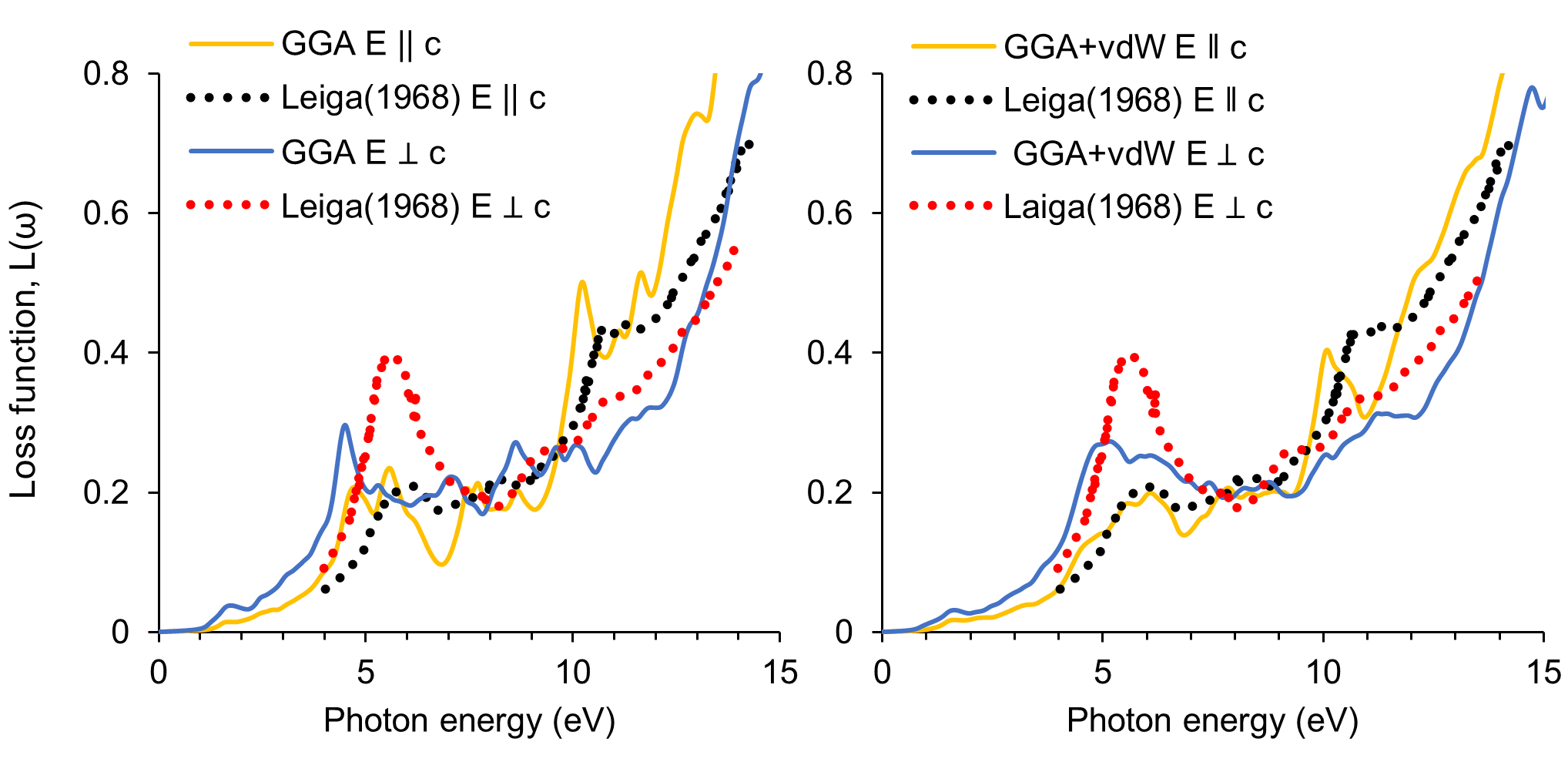}
    \caption{Using data taken by Leiga (1968) \cite{Leiga:68, Leiga:682} we can compare the calculations for bulk properties of the Energy Loss made using GGA (Left) and GGA+vdW (Right) to experimental data taken when the photon is polarized parallel and perpendicular to the crystal.}
    \label{fig:EnergyLossComparison}
\end{figure}

Similar comparisons for the index of refraction and the loss function follow with subsequent comparisons to data in Figure \ref{fig:EnergyLossComparison} having generally the same conclusions as those given above.

\begin{figure}[htb]
    \centering
    \includegraphics[scale=0.26]{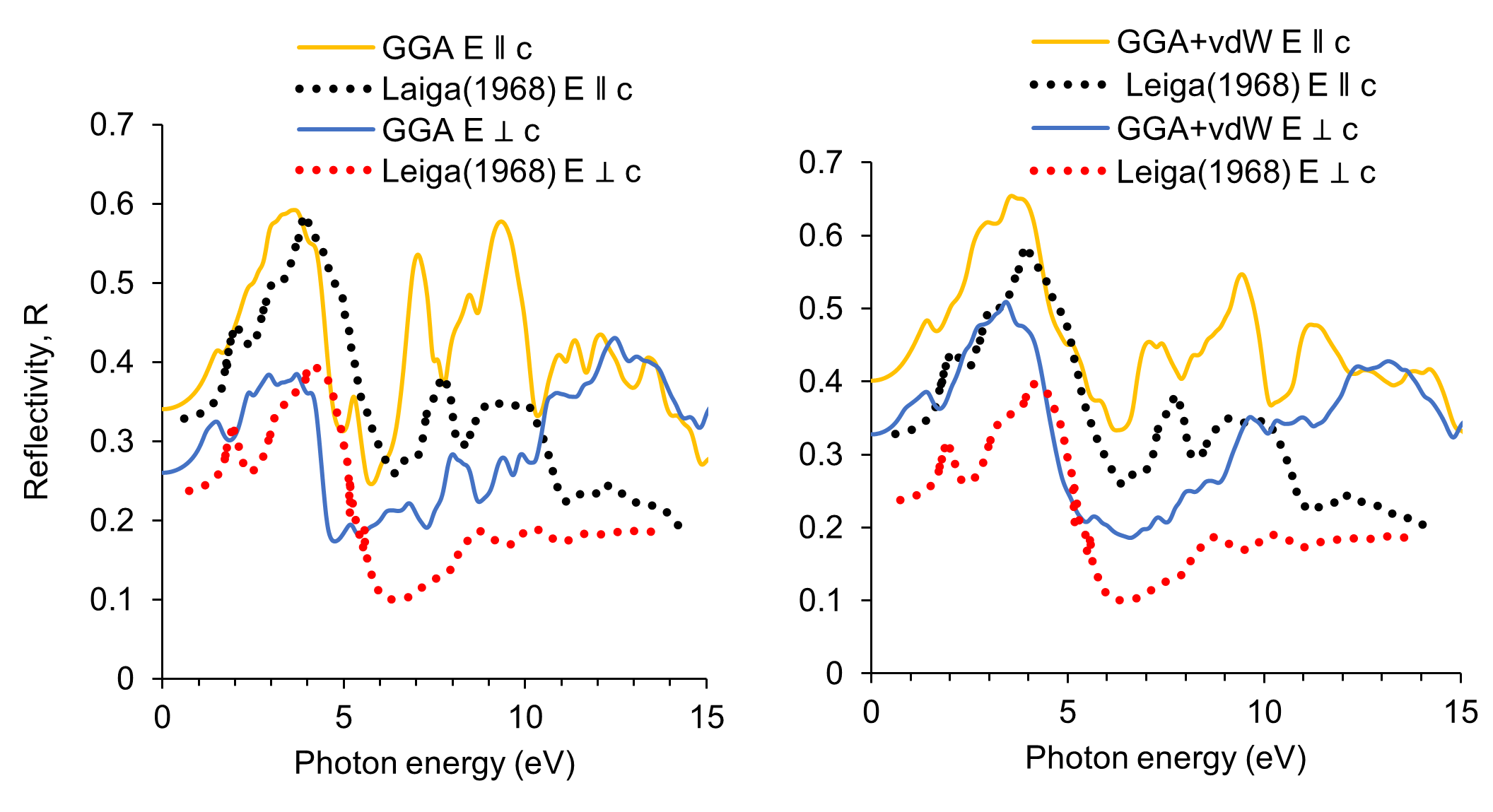}
    \caption{Using data taken by Leiga (1968) \cite{Leiga:68, Leiga:682} we can compare the calculations for bulk properties of the reflectance made using GGA (Left) and GGA+vdW (Right) to experimental data taken when the photon is polarized parallel and perpendicular to the crystal. Broad spectral properties show general agreement with the shape of the spectrum having approximate agreement but the magnitude of the reflectance differing between $\sim 25-50\%$ across the spectrum.}
    \label{fig:ReflectanceComparison}
\end{figure}


\FloatBarrier

\section{Conclusion}\label{conclusion}
In this work, we have employed density functional theory to systematically investigate the thermodynamic, electronic and optical properties of $(10\overline{1}0)$ surface layers of trigonal selenium (\textit{t}-Se). We find that within few atomic layers of \textit{t}-Se, free standing thin films started to have bulk like properties. From the electronic structures calculations, both the valence and the conduction bands' extrema (VBM and CBM) have Se-$p$ characters, so optical absorption across the band gap is highly suppressed. In addition, an indirect to direct band gap transition was observed as the number of layers increases within the slab. Overall, the optical properties matches well with the experimental results when Van der Waals corrections are included in the DFT Hamiltonian. The optical absorption peaks near the area of our interest ($\sim$9 eV) and thus should have very favorable properties sought in their application as a VUV photodetector. 

This work serves as a foundational starting point for theoretical calculations to better understand the properties a selenium based detector will have for photons in the VUV spectrum. The calculations have shown the surface energy, work function, electronic, and optical properties which can be derived for \textit{t}-Se. Where possible, the calculations are compared to existing data. The general agreement found between the existing data for \textit{t}-Se and the work done here opens up the doorway to further analysis of the amorphous state (a-Se), which is most commonly used in photon detectors and helps better understand the fundamental properties of the thin film selenium detectors which can be employed for VUV photon detection given Se's favorable properties of high absorption coefficient, low loss function, and favorable transport properties.

\section*{Acknowledgments}
All computations for this work were executed in Texas Advanced Computing Center (TACC) at Austin, TX. This work was supported by a grant from the U.S. Department of Energy, Office of Science, Office of High Energy Physics under Award Number DE-SC 0000253485.

\section*{Conflict of Interest and Data Sharing}
The authors declare no conflict of interest in this work. The data that support the findings of this study are available from the corresponding author upon reasonable request.

\bibliographystyle{JHEP}
\bibliography{bibliography}

\providecommand{\href}[2]{#2}\begingroup\raggedright\begin{thebibliography}{10}

\bibitem{OldSelenium}
{\it Effect of light on selenium during the passage of an electric current},
  {\em Nature} {\bf 7} (Feb, 1873) 303--303.

\bibitem{OldSelenium2}
W.~Smith, {\it The action of light on selenium},  {\em Journal of the Society
  of Telegraph Engineers} {\bf 2} (Feb, 1873) 31--33.

\bibitem{10.1579/0044-7447(2007)36[94:SGAH]2.0.CO;2}
F.~Fordyce, {\it {Selenium Geochemistry and Health}},  {\em AMBIO: A Journal of
  the Human Environment} {\bf 36} (2007), no.~1 94 -- 97.

\bibitem{PhysRev.158.623}
S.~Tutihasi and I.~Chen, {\it Optical properties and band structure of trigonal
  selenium},  {\em Phys. Rev.} {\bf 158} (Jun, 1967) 623--630.

\bibitem{doi:10.1063/1.1702669}
H.~T. Li and P.~J. Regensburger, {\it Photoinduced discharge characteristics of
  amorphous selenium plates},  {\em Journal of Applied Physics} {\bf 34}
  (1963), no.~6 1730--1735,
  [\href{http://xxx.lanl.gov/abs/https://doi.org/10.1063/1.1702669}{{\tt
  https://doi.org/10.1063/1.1702669}}].

\bibitem{IEEEexample:Xerox}
R.~P. Sechak, {\it Tri-layer selenium doped photoreceptor},  1972.

\bibitem{doi:10.1063/1.881994}
J.~Rowlands and S.~Kasap, {\it Amorphous semiconductors usher in digital
  x‐ray imaging},  {\em Physics Today} {\bf 50} no.~11 24.

\bibitem{doi.org/10.1023/A:1008993813689}
S.~Kasap and J.~Rowlands, {\it Review x-ray photoconductors and stabilized a-se
  for direct conversion digital flat-panel x-ray image-detectors},  {\em
  Journal of Materials Science: Materials in Electronics} {\bf 11} (2000),
  no.~6 179–198,
  [\href{http://xxx.lanl.gov/abs/https://doi.org/10.1023/A:1008993813689}{{\tt
  https://doi.org/10.1023/A:1008993813689}}].

\bibitem{Masuzawa_2013}
T.~Masuzawa, I.~Saito, T.~Yamada, M.~Onishi, H.~Yamaguchi, Y.~Suzuki,
  K.~Oonuki, N.~Kato, S.~Ogawa, Y.~Takakuwa, and et~al., {\it Development of an
  amorphous selenium-based photodetector driven by a diamond cold cathode},
  {\em Sensors} {\bf 13} (Oct, 2013) 13744–13778.

\bibitem{doi:10.1118/1.2008428}
W.~Zhao, D.~Li, A.~Reznik, B.~J.~M. Lui, D.~C. Hunt, J.~A. Rowlands, Y.~Ohkawa,
  and K.~Tanioka, {\it Indirect flat-panel detector with avalanche gain:
  Fundamental feasibility investigation for sharp-amfpi (scintillator harp
  active matrix flat panel imager)},  {\em Medical Physics} {\bf 32} (2005),
  no.~9 2954--2966,
  [\href{http://xxx.lanl.gov/abs/https://aapm.onlinelibrary.wiley.com/doi/pdf/10.1118/1.2008428}{{\tt
  https://aapm.onlinelibrary.wiley.com/doi/pdf/10.1118/1.2008428}}].

\bibitem{Nakada_1985}
T.~Nakada and A.~Kunioka, {\it Polycrystalline thin-film {TiO}2/se solar
  cells},  {\em Japanese Journal of Applied Physics} {\bf 24} (jul, 1985)
  L536--L538.

\bibitem{Ito_1982}
H.~Ito, M.~Oka, T.~Ogino, A.~Takeda, and Y.~Mizushima, {\it Selenium thin film
  solar cell},  {\em Japanese Journal of Applied Physics} {\bf 21} (jan, 1982)
  77.

\bibitem{Todorov2017}
T.~K. Todorov, S.~Singh, and et. al, {\it Ultrathin high band gap solar cells
  with improved efficiencies from the world’s oldest photovoltaic material},
  {\em Nature Communications} {\bf 8} (Sept, 2017).

\bibitem{https://doi.org/10.1002/pssr.201370438}
T.~Masuzawa, M.~Onishi, I.~Saito, T.~Yamada, A.~T.~T. Koh, D.~H.~C. Chua,
  S.~Ogawa, Y.~Takakuwa, Y.~Mori, T.~Shimosawa, and K.~Okano, {\it Cover
  picture: High quantum efficiency uv detection using a-se based photodetector
  (phys. status solidi rrl 7/2013)},  {\em physica status solidi (RRL) –
  Rapid Research Letters} {\bf 7} (2013), no.~7
  [\href{http://xxx.lanl.gov/abs/https://onlinelibrary.wiley.com/doi/pdf/10.1002/pssr.201370438}{{\tt
  https://onlinelibrary.wiley.com/doi/pdf/10.1002/pssr.201370438}}].

\bibitem{6399582}
S.~{Abbaszadeh}, K.~S. {Karim}, and V.~{Karanassios}, {\it Measurement of uv
  from a microplasma by a microfabricated amorphous selenium detector},  {\em
  IEEE Transactions on Electron Devices} {\bf 60} (2013), no.~2 880--883.

\bibitem{C8TC05873C}
M.~Zhu, G.~Niu, and J.~Tang, {\it Elemental se: fundamentals and its
  optoelectronic applications},  {\em J. Mater. Chem. C} {\bf 7} (2019)
  2199--2206.

\bibitem{app11062455}
K.~Majumdar and K.~Mavrokoridis, {\it Review of liquid argon detector
  technologies in the neutrino sector},  {\em Applied Sciences} {\bf 11}
  (2021), no.~6.

\bibitem{BAUDIS201450}
L.~Baudis, {\it Wimp dark matter direct-detection searches in noble gases},
  {\em Physics of the Dark Universe} {\bf 4} (2014) 50--59. DARK TAUP2013.

\bibitem{instruments5010004}
M.~Kuźniak and A.~M. Szelc, {\it Wavelength shifters for applications in
  liquid argon detectors},  {\em Instruments} {\bf 5} (2021), no.~1.

\bibitem{Ieki:2018pbf}
K.~Ieki et~al., {\it {Large-Area MPPC with Enhanced VUV Sensitivity for Liquid
  Xenon Scintillation Detector}},  {\em Nucl. Instrum. Meth. A} {\bf 925}
  (2019) 148--155, [\href{http://xxx.lanl.gov/abs/1809.08701}{{\tt
  arXiv:1809.08701}}].

\bibitem{Asaadi:2018ixs}
J.~Asaadi, B.~J.~P. Jones, A.~Tripathi, I.~Parmaksiz, H.~Sullivan, and Z.~G.~R.
  Williams, {\it {Emanation and bulk fluorescence in liquid argon from
  tetraphenyl butadiene wavelength shifting coatings}},  {\em JINST} {\bf 14}
  (2019), no.~02 P02021, [\href{http://xxx.lanl.gov/abs/1804.00011}{{\tt
  arXiv:1804.00011}}].

\bibitem{Abraham:2021otn}
Y.~Abraham et~al., {\it {Wavelength-Shifting Performance of Polyethylene
  Naphthalate Films in a Liquid Argon Environment}},
  \href{http://xxx.lanl.gov/abs/2103.03232}{{\tt arXiv:2103.03232}}.

\bibitem{OOTANI2015220}
W.~Ootani, K.~Ieki, T.~Iwamoto, D.~Kaneko, T.~Mori, S.~Nakaura, M.~Nishimura,
  S.~Ogawa, R.~Sawada, N.~Shibata, Y.~Uchiyama, K.~Yoshida, K.~Sato, and
  R.~Yamada, {\it Development of deep-uv sensitive mppc for liquid xenon
  scintillation detector},  {\em Nuclear Instruments and Methods in Physics
  Research Section A: Accelerators, Spectrometers, Detectors and Associated
  Equipment} {\bf 787} (2015) 220--223. New Developments in Photodetection
  NDIP14.

\bibitem{Leiga:68}
A.~G. Leiga, {\it Optical properties of trigonal selenium in the vacuum
  ultraviolet$\ast$},  {\em J. Opt. Soc. Am.} {\bf 58} (Jul, 1968) 880--884.

\bibitem{Leiga:682}
A.~G. Leiga, {\it Optical properties of amorphous selenium in the vacuum
  ultraviolet},  {\em J. Opt. Soc. Am.} {\bf 58} (Nov, 1968) 1441--1445.

\bibitem{Hohenberg1964}
P.~Hohenberg and W.~Kohn, {\it Inhomogeneous electron gas},  {\em Phys. Rev.}
  {\bf 136} (Nov, 1964) B864--B871.

\bibitem{Kohn1965}
W.~Kohn and L.~J. Sham, {\it {Self-Consistent Equations Including Exchange and
  Correlation Effects}},  {\em Physical Review} {\bf 140} (nov, 1965)
  A1133--A1138, [\href{http://xxx.lanl.gov/abs/PhysRev.140.A1133}{{\tt
  PhysRev.140.A1133}}].

\bibitem{PAW-PhysRevB.50.17953}
P.~E. Bl\"ochl, {\it Projector augmented-wave method},  {\em Phys. Rev. B} {\bf
  50} (Dec, 1994) 17953--17979.

\bibitem{Kresse-PAW-PhysRevB.59.1758}
G.~Kresse and D.~Joubert, {\it From ultrasoft pseudopotentials to the projector
  augmented-wave method},  {\em Phys. Rev. B} {\bf 59} (Jan, 1999) 1758--1775.

\bibitem{KRESSE199615}
G.~Kresse and J.~Furthmüller, {\it Efficiency of ab-initio total energy
  calculations for metals and semiconductors using a plane-wave basis set},
  {\em Computational Materials Science} {\bf 6} (1996), no.~1 15 -- 50.

\bibitem{Kresse-PhysRevB.54.11169}
G.~Kresse and J.~Furthm\"uller, {\it Efficient iterative schemes for ab initio
  total-energy calculations using a plane-wave basis set},  {\em Phys. Rev. B}
  {\bf 54} (Oct, 1996) 11169--11186.

\bibitem{vesta-Momma:db5098}
K.~Momma and F.~Izumi, {\it {{\it VESTA3} for three-dimensional visualization
  of crystal, volumetric and morphology data}},  {\em Journal of Applied
  Crystallography} {\bf 44} (Dec, 2011) 1272--1276.

\bibitem{GGA-PBE-PhysRevLett.77.3865}
J.~P. Perdew, K.~Burke, and M.~Ernzerhof, {\it Generalized gradient
  approximation made simple},  {\em Phys. Rev. Lett.} {\bf 77} (Oct, 1996)
  3865--3868.

\bibitem{GGA-PBE-errata-PhysRevLett.78.1396}
J.~P. Perdew, K.~Burke, and M.~Ernzerhof, {\it Generalized gradient
  approximation made simple [phys. rev. lett. 77, 3865 (1996)]},  {\em Phys.
  Rev. Lett.} {\bf 78} (Feb, 1997) 1396--1396.

\bibitem{Grimme-D3-doi:10.1063/1.3382344}
S.~Grimme, J.~Antony, S.~Ehrlich, and H.~Krieg, {\it A consistent and accurate
  ab initio parametrization of density functional dispersion correction (dft-d)
  for the 94 elements h-pu},  {\em The Journal of Chemical Physics} {\bf 132}
  (2010), no.~15 154104,
  [\href{http://xxx.lanl.gov/abs/https://doi.org/10.1063/1.3382344}{{\tt
  https://doi.org/10.1063/1.3382344}}].

\bibitem{doi:10.1063/1.3382344}
S.~Grimme, J.~Antony, S.~Ehrlich, and H.~Krieg, {\it A consistent and accurate
  ab initio parametrization of density functional dispersion correction (dft-d)
  for the 94 elements h-pu},  {\em The Journal of Chemical Physics} {\bf 132}
  (2010), no.~15 154104,
  [\href{http://xxx.lanl.gov/abs/https://doi.org/10.1063/1.3382344}{{\tt
  https://doi.org/10.1063/1.3382344}}].

\bibitem{Wyckorff}
R.~Wyckoff, {\it Crystal structures, selenium}, .

\bibitem{doi:10.1021/ic50054a037}
P.~Cherin and P.~Unger, {\it The crystal structure of trigonal selenium},  {\em
  Inorganic Chemistry} {\bf 6} (1967), no.~8 1589--1591,
  [\href{http://xxx.lanl.gov/abs/https://doi.org/10.1021/ic50054a037}{{\tt
  https://doi.org/10.1021/ic50054a037}}].

\bibitem{Cohesive-PhysRevB.27.6296}
D.~Vanderbilt and J.~D. Joannopoulos, {\it Total energies in se. i. the
  trigonal crystal},  {\em Phys. Rev. B} {\bf 27} (May, 1983) 6296--6301.

\bibitem{cohesive-si-PhysRevB.43.14248}
B.~Farid and R.~W. Godby, {\it Cohesive energies of crystals},  {\em Phys. Rev.
  B} {\bf 43} (Jun, 1991) 14248--14250.

\bibitem{Boettger}
J.~C. Boettger, {\it Nonconvergence of surface energies obtained from thin-film
  calculations},  {\em Phys. Rev. B} {\bf 49} (Jun, 1994) 16798--16800.

\bibitem{sajib-ga2o3-https://doi.org/10.1002/pssr.201800554}
S.~K. Barman and M.~N. Huda, {\it Mechanism behind the easy exfoliation of
  ga2o3 ultra-thin film along (100) surface},  {\em physica status solidi (RRL)
  – Rapid Research Letters} {\bf 13} (2019), no.~5 1800554,
  [\href{http://xxx.lanl.gov/abs/https://onlinelibrary.wiley.com/doi/pdf/10.1002/pssr.201800554}{{\tt
  https://onlinelibrary.wiley.com/doi/pdf/10.1002/pssr.201800554}}].

\bibitem{shafaq-MOTEN201637}
S.~A. Moten, R.~Atta-Fynn, A.~K. Ray, and M.~N. Huda, {\it Size effects on the
  electronic and magnetic properties of puo2 (111) surface},  {\em Journal of
  Nuclear Materials} {\bf 468} (2016) 37 -- 45.

\bibitem{edan-BAINGLASS2021121762}
E.~Bainglass and M.~N. Huda, {\it Low-index stoichiometric surfaces of
  cubiw2o8},  {\em Surface Science} {\bf 705} (2021) 121762.

\bibitem{ALKHALDI2019e02908}
N.~D. Alkhaldi, S.~K. Barman, and M.~N. Huda, {\it Crystal structures and the
  electronic properties of silicon-rich silicon carbide materials by first
  principle calculations},  {\em Heliyon} {\bf 5} (2019), no.~11 e02908.

\bibitem{LEE1971213}
L.-H. Lee, {\it Solid surface tensions of amorphous and crystalline selenium},
  {\em Journal of Non-Crystalline Solids} {\bf 6} (1971), no.~3 213--220.

\bibitem{Guisbeirs-doi:10.1063/1.4769358}
G.~Guisbiers, S.~Arscott, and R.~Snyders, {\it An accurate determination of the
  surface energy of solid selenium},  {\em Applied Physics Letters} {\bf 101}
  (2012), no.~23 231606,
  [\href{http://xxx.lanl.gov/abs/https://doi.org/10.1063/1.4769358}{{\tt
  https://doi.org/10.1063/1.4769358}}].

\bibitem{underestimation-gap-https://doi.org/10.1002/qua.560280846}
J.~P. Perdew, {\it Density functional theory and the band gap problem},  {\em
  International Journal of Quantum Chemistry} {\bf 28} (1985), no.~S19
  497--523,
  [\href{http://xxx.lanl.gov/abs/https://onlinelibrary.wiley.com/doi/pdf/10.1002/qua.560280846}{{\tt
  https://onlinelibrary.wiley.com/doi/pdf/10.1002/qua.560280846}}].

\bibitem{underestimation-gap2-Perdew2801}
J.~P. Perdew, W.~Yang, K.~Burke, Z.~Yang, E.~K.~U. Gross, M.~Scheffler, G.~E.
  Scuseria, T.~M. Henderson, I.~Y. Zhang, A.~Ruzsinszky, H.~Peng, J.~Sun,
  E.~Trushin, and A.~G{\"o}rling, {\it Understanding band gaps of solids in
  generalized kohn{\textendash}sham theory},  {\em Proceedings of the National
  Academy of Sciences} {\bf 114} (2017), no.~11 2801--2806,
  [\href{http://xxx.lanl.gov/abs/https://www.pnas.org/content/114/11/2801.full.pdf}{{\tt
  https://www.pnas.org/content/114/11/2801.full.pdf}}].

\end{thebibliography}\endgroup


\providecommand{\href}[2]{#2}\begingroup\raggedright\endgroup
\newpage
\section{Supplemental Material}

\vspace{10mm}
\begin{center}
\Large{First principles studies of the surface and opto-electronic properties of ultra-thin \textit{t}-Se}\\
\vspace{2mm}
\large{S. K. Barman, M. N. Huda, J. Asaadi, E. Gramellini, D. Nygren}
\end{center}

\vspace{10mm}



\subsection{Surface Properties}\label{sec:Supp-SurfaceCalc}


\begin{table}[!htb]
    \centering
    \begin{tabular}{|c|c|c|}
    \hline
    & \multicolumn{2}{c|}{Surface area of free-standing laryers (\r{A}$^2$)} \\
    \hline
    Number of layers & \textbf{GGA} & \textbf{GGA + vdW} \\
    \hline
    1 & 21.08 & 20.29 \\
    \hline
    2 & 21.91 & 20.84 \\
    \hline
    3 & 22.23 & 21.08 \\
    \hline
    4 & 22.36 & 21.17 \\
    \hline
    5 & 22.44 & 21.22 \\
    \hline
    6 & 22.52 & 21.32 \\
    \hline
    7 & 22.62 & 21.35 \\
    \hline
    8 & 22.64 & 21.41 \\
    \hline
    9 & 22.61 & 21.43 \\
    \hline
    10 & 22.64 & 21.46 \\
    \hline
    11 & 22.77 & 21.46 \\
    \hline
    12 & 22.71 & 21.46 \\
    \hline
    13 & 22.68 & 21.44 \\
    \hline
    14 & 22.73 & 21.51 \\
    \hline
    15 & 22.74 & 21.53 \\
    \hline
    \end{tabular}
    \caption{GGA and GGA + vdW relaxed $(10\overline{1}0)$ surface areas of \textit{t}-Se as the number of layer increases for free standing layers.}
    \label{tab:SurfaceAreas}
\end{table}

\begin{table}[!htb]
    \centering
    \begin{tabular}{|c|c|c|}
    \hline
    & \multicolumn{2}{c|}{Surface energies, $\gamma\text{(mJ/m}^2$) for $(10\overline{1}0)$ surface slabs of \textit{t}-Se} \\
    \hline
    Number of layers in slabs & GGA & GGA \\ 
    \hline
    1 & 160.22 & 176.24 \\ 
    \hline
    2 & 160.22 & 192.26 \\
    \hline
    3 & 160.22 & 192.26 \\
    \hline
    4 & 160.22 & 192.26 \\
    \hline
    5 & 160.22 & 192.26 \\
    \hline
    6 & 160.22 & 192.26 \\
    \hline
    7 & 160.22 & 192.26 \\
    \hline
    8 & 160.22 & 192.26 \\
    \hline
    9 & 160.22 & 192.26 \\
    \hline
    10& 160.22 & 192.26 \\
    \hline
    11& 160.22 & 192.26 \\
    \hline
    12& 160.22 & 176.24 \\
    \hline
    13& 160.22 & 176.24 \\
    \hline
    14& 160.22 & 176.24 \\
    \hline
    15& 160.22 & 176.24 \\
    \hline
    \end{tabular}
    \caption{GGA and GGA + vdW calculated energies of fully relaxed $(10\overline{1}0)$ surface slabs as a function of the number of layers. Atoms in this relaxation (free-standing layers) could relax in all three directions inside the simulation box.}
    \label{tab:TotalEnergyByLayer}
\end{table}

\begin{table}[!htb]
    \centering
    \begin{tabular}{|c|c|c|}
    \hline
    &\multicolumn{2}{c|}{Slab Thickness (\r{A})} \\
    \hline
    Number of layers in slabs & GGA & GGA + vdW \\
    \hline
    2 & 4.583 & 4.283 \\
    \hline
    3 & 9.056 & 8.531 \\
    \hline
    4 & 14.592 & 12.690 \\
    \hline
    5 & 18.065 & 16.903 \\
    \hline
    6 & 22.628 & 21.123 \\
    \hline
    7 & 27.150 & 29.614 \\
    \hline
    8 & 31.568 & 29.614 \\
    \hline
    9 & 36.109 & 33.788 \\
    \hline
    10& 40.604 & 38.081 \\
    \hline
    11& 45.087 & 42.232 \\
    \hline
    12& 49.604 & 46.427 \\
    \hline
    13& 54.102 & 50.576 \\
    \hline
    14& 58.617 & 54.919 \\
    \hline
    15& 63.045 & 59.217 \\
    \hline
    \end{tabular}
    \caption{GGA and GGA + vdW optimized slab thickness as a function of the number of layers.}
    \label{tab:LayerThickness}
\end{table}

\FloatBarrier
\newpage

\subsection{Bulk Electronic Properties}\label{sec:Supp-Bulk-Elec-Prop}

\begin{figure}[htb]
    \centering
    \includegraphics[scale=0.4]{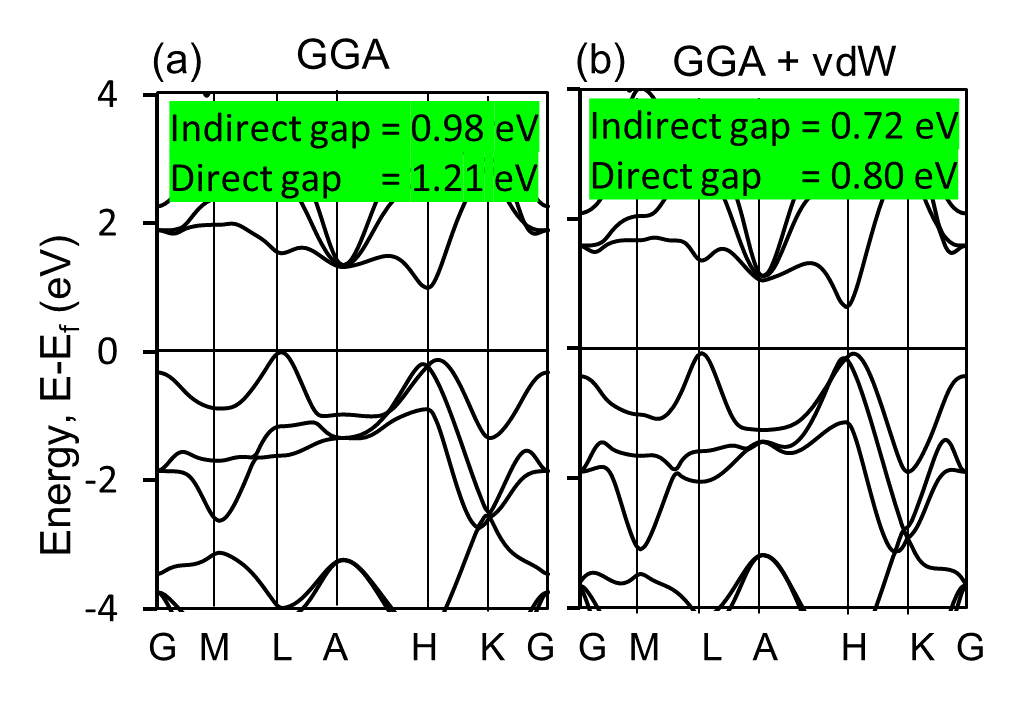}
    \caption{(a) GGA and (b) GGA + vdW calculated band structures of pure bulk \textit{t}-Se. The calculated indirect(direct) band gaps are 0.98(1.21) $eV$ and 0.72(0.80) $eV$, respectively. The Fermi level, $E_F$ is set at 0 $eV$. van der Waals correction, vdW was applied through the zero damping DFT-D3 method of Grimme.}
    \label{fig:BandStructure-Bulk}
\end{figure}

\begin{figure}[htb]
    \centering
    \includegraphics[scale=0.6]{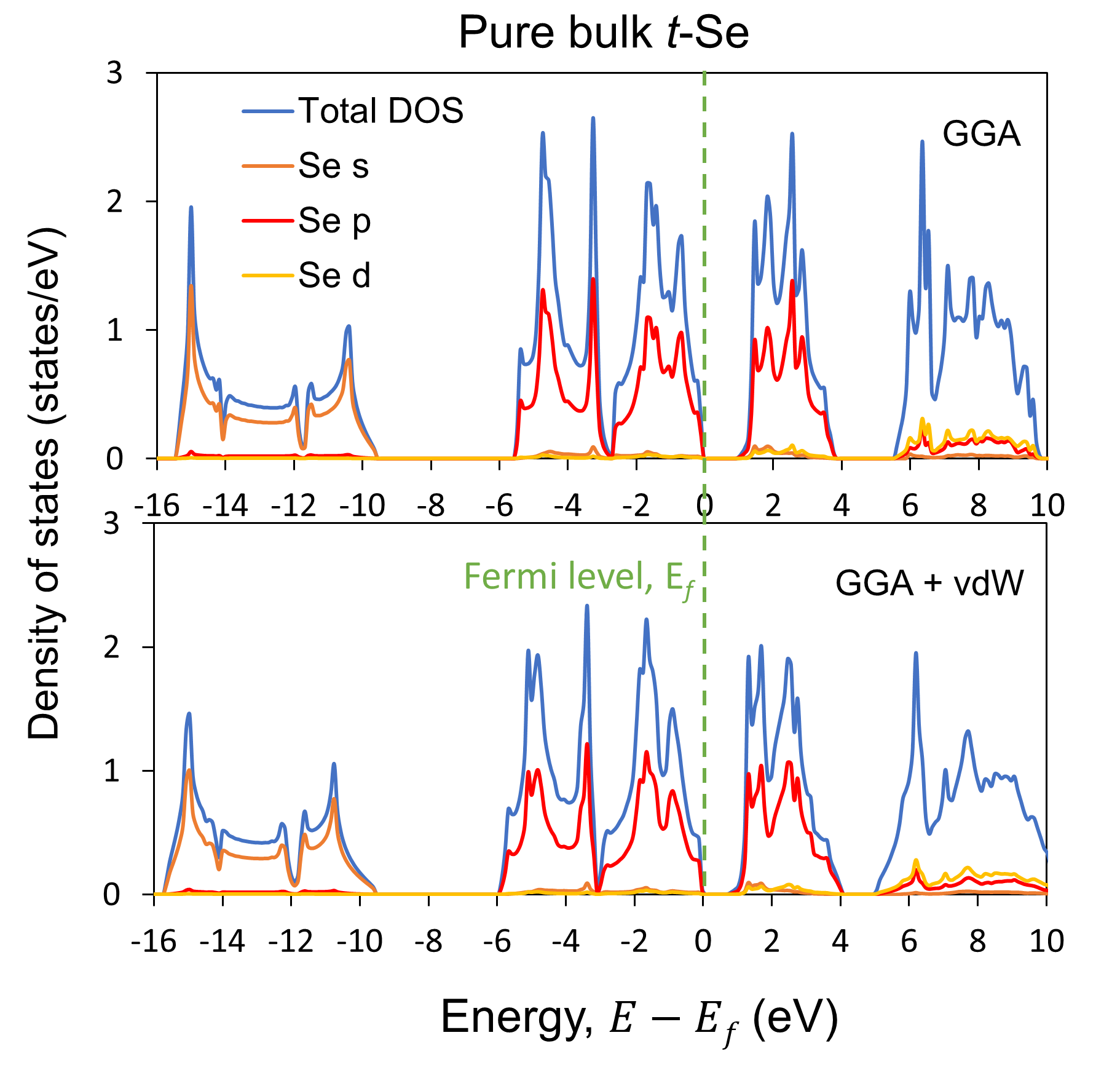}
    \caption{GGA and GGA + vdW calculated the density of states (DOS) of pure bulk \textit{t}-Se. This figure shows that the top of the valence band and the bottom of the conduction band are mainly composed of Se $p$. Hence, the bands in Figure \ref{fig:BandStructure-Bulk} in the energy range of (-4, 4) are contributed primarily by Se $p$. The Fermi level, $E_F$ in this figure is set at 0 $eV$.}
    \label{fig:DOS-Bulk}
\end{figure}

\FloatBarrier



\newpage



\end{document}